\documentclass[useAMS,usenatbib]{mn2e}

\usepackage{graphicx}

\title[MC Simulations of the Broadband Spectra of Sagittarius A* through the use of GRMHD]{Monte Carlo Simulations of the Broadband Spectra of Sagittarius A* through the use of General Relativistic MHD}
\author[Guy Hilburn, Edison Liang, Siming Liu, and Hui Li]{Guy Hilburn$^{1,2}$, Edison Liang$^{1}$
, Siming Liu$^{3}$, and Hui Li$^{2}$\\
$^{1}$Department of Physics and Astronomy, Rice University, 6100 Main Street, Houston, Texas, 77005\\
$^{2}$Los Alamos National Laboratory, P.O. Box 1663, Los Alamos, New Mexico, 87545\\
$^{3}$Department of Physics and Astronomy, University of Glasgow, G12 8QQ, Scotland\\}
\begin{document}

\date{}

\pagerange{\pageref{firstpage}--\pageref{lastpage}} \pubyear{2002}

\label{firstpage}

\maketitle

\begin{center}
\begin{abstract}
We present results of simulations of the spectrum of the accretion flow onto the supermassive black hole in our Galactic Centre, Sagittarius A*, generated with a coupling of Monte-Carlo (MC) radiation and general relativistic magnetohydrodynamic (GRMHD) codes.  In our modeling, we use the 2D HARM GRMHD code to first model the physical parameters of the disk, then feed its results into our 2D MC photon transport code.  We will discuss results obtained which fit radio, IR, and Chandra-obtained flaring or quiescent x-ray data points, as well as the validity of the amount of scaling of input parameters (density, temperature, and magnetic field) required to fit these points.  HARM output will be used to suggest whether the scaling is within reasonable limits.
\end{abstract}
\end{center}


\section{Introduction}

In recent years, a great deal of effort has been made to understand the complete picture of Sagittarius A* (Sgr A*), a radio/IR/x-ray emission source at our Galactic Centre.  It is widely accepted that this source is related to the accretion flow of a supermassive black hole whose mass we have taken to be 3.6 million solar masses \citep{b11,b20,b15}.

The spectrum of Sgr A* shows several important components or signatures.  \citet{b4} first discovered the source in the radio/NIR, and ensuing observations by a number of researchers confirmed the strength of the source to be primarily in these regimes.  Years later, Sgr A*'s x-ray spectrum was explored by \citet{b2,b3}, which found notable signatures higher than 10$^{17}$ Hz, and also found a variability in this range, which suggests separate spectral states -- flaring and quiescent.  Simultaneous multi-wavelength observations have generally supported this suggestion \citep{b35,b36,b37}.

A number of models have arisen to explain the mechanisms at work to produce the spectra of black holes.  The standard-disk idea was first explored by \citet{b24} and proposes a situation in which gravitational energy is efficiently converted to radiation.  This is most appropriate for optically thick disks with the flows on nearly Keplerian orbits.  Sgr A*'s luminosity is less than 10$^{-8}$ L$_{Edd}$, where the Eddington luminosity L$_{Edd}$ represents the luminosity level where the gravitational force inward equals the radiation force outward for spherical accretions.  Sources with such low luminosities may be described by the standard-disk idea in cases where the disk remains cold and is optically thick, and the luminosity is low primarily due to a very low accretion rate.  However, the accretion flow in Sgr A* must be very hot, and there is no evidence for a cold, optically thick disk component.  The low luminosity is limited by both a low accretion rate and a low emission efficiency of the flow.  In cases such as this, other accretion models need to be considered.

One developed model is called the advection dominated accretion flow (ADAF) \citep{b25}.  This suggested that close to the horizon, much of the energy of the accretion flow was advected into the black hole, rather than being radiated away, and was explored by a number of groups, including \citet{b16}, \citet{b1}, and \citet{b21}.  It was found that the ADAF solution can fit Sgr A* data well \citep{b26}, but while promising, this approach had several major drawbacks, namely its one dimensional approach and simplification of magnetic fields.  The X-ray emission is produced primarily at large radii, which cannot account for the observed short-timescale X-ray flares.  In an attempt to consider an alternative approach, \citet{b21} suggested inclusion of nonthermal electrons and found satisfactory fits, but again with simple magnetic fields.  

It became apparent that a more complicated treatment of magnetic fields would be important for a more accurate simulation, as the magnetorotational instability was found to be of vital importance in the development of turbulence that drives the accretion flow \citep{b12,b27,b28}.  This instability ensures that in an accretion disk environment the necessity of outward angular momentum transport leads to the establishment of very complex flow and field patterns.

Models have since begun using magnetohydrodynamic codes to simulate accretion flows, with some measure of success.  \citet{b18} ran emission simulations on an MHD model by \citet{b23} and was able to fit data well, but had difficulties doing so without cutting out a large portion of their simulation volume.  \citet{b29} calculated the radio spectrum based on MHD simulations done by \citet{b30}, but not other spectral bands.

We hope to show a simulation method which minimizes compromises on consistency via assumptions or simplifications by using the best tools available to present a more realistic accretion disk picture.

We will present our approach to Sgr A* modeling using GRMHD and MC methods in \S 2, present the results gained through these methods in \S 3, and revisit the most important revelations and conclusions made within in \S 4.

\section{Simulation Method}

Our modeling method involves the coupling of two very different codes which complement each other to provide a consistent view of the physical and spectral conditions in the accretion disk of Sagittarius A*.  Detailed in the appropriate sections below are descriptions of the codes, discussions of their appropriateness for our work, and the manner in which they were coupled to provide an overall model.

\subsection{HARM GRMHD physical space modeling}

The GRMHD code selected to determine the model's physical space is the 2D HARM code presented and detailed in \citet{b10} and \citet{b17}.  It is not within the scope of this paper to completely detail the HARM code's inner workings, so readers should see these referenced papers for further information.

The code solves hyperbolic partial differential equations in conservative form -- uniquely suiting it to a number of astrophysical problems, specifically those involving magnetohydrodynamics in areas where general relativity is important.  As the code evolves the space through time, conserved variables are converted to primitive variables at each step, to calculate a set of fluxes, given a set of sources.  Use of primitive variable calculations allows the code to work with analytic solutions, rather than finding solutions numerically -- leading to much faster calculation time.

In this case, the variables tracked include density, total energy, internal energy, flow velocities, and magnetic fields.  The last two are calculated both as 3-component and 4-component tensor quantities.

\subsubsection{Model set-up for our work}

As imported, HARM was set-up quite appropriately for our purposes.  It was configured to evolve an accretion disk about a black hole, given a number of user-controlled parameters, on a 2D grid spaced (in sperical coordinates) radially and angularly, and assumed to be axially symmetric about the black hole's spin axis.  The cells are assumed to be uniformly spaced with regard to a set of coordinates X1 and X2, which can be converted then to r and theta, respectively.  This conversion leads to a logarithmic spacing in r, with cells closer to the horizon having greater resolution (and smaller size), and more concentrated cells closer to the equator.  This effectively increases resolution in the plane of the accretion disk and close to the black hole, where the detail is most useful due to the much smaller length scales of interest in these regions.  The resolution increase toward the horizon is vital to the successful running of the code, as it helps maintain proper cell aspect ratios on the polar grid.  A graphical representation of HARM's grid layout can be seen in Figure 5, where it is compared to the MC code's grid.

As the start, HARM seeds an equilibrium torus around the black hole with density as detailed by \citet{b9}.  The torus is perturbed by adding a small poloidal magnetic field and allowed to evolve around the black hole.  We have not attempted runs with an initial toroidal, azimuthal field, though this is allowed by axial symmetry, and should serve to increase the total field at the end of simulation.  The additional toroidal component should not influence the MRI development, instead only adding to the final field.

While most default parameters as included are appropriate for the simulation, several had to be tweaked for this project.  To effectively simulate conditions near an a low luminosity Active Galactic Nuclei, where the nonrelativistic gas pressure presumably dominates, an adiabatic index of 5/3 was chosen.  As a first approximation, we have chosen a non-spinning black hole.  Future work will study the effects of including black hole spin.  The simulation volume ranges from just inside the horizon at 2 GM/c$^{2}$ to 40 GM/c$^{2}$ -- with the initial torus having an inner radius of 6 GM/c$^{2}$ and its pressure maximum at 14.7 GM/c$^{2}$.  Our final trial was done on a grid of 512x512 -- that is, 512 radial cells by 512 angular cells.  Output from the code is all scaled to M, the black hole mass, for near complete freedom.

\subsubsection{Results and interpretations}

\begin{figure}
  \includegraphics[scale=0.22]{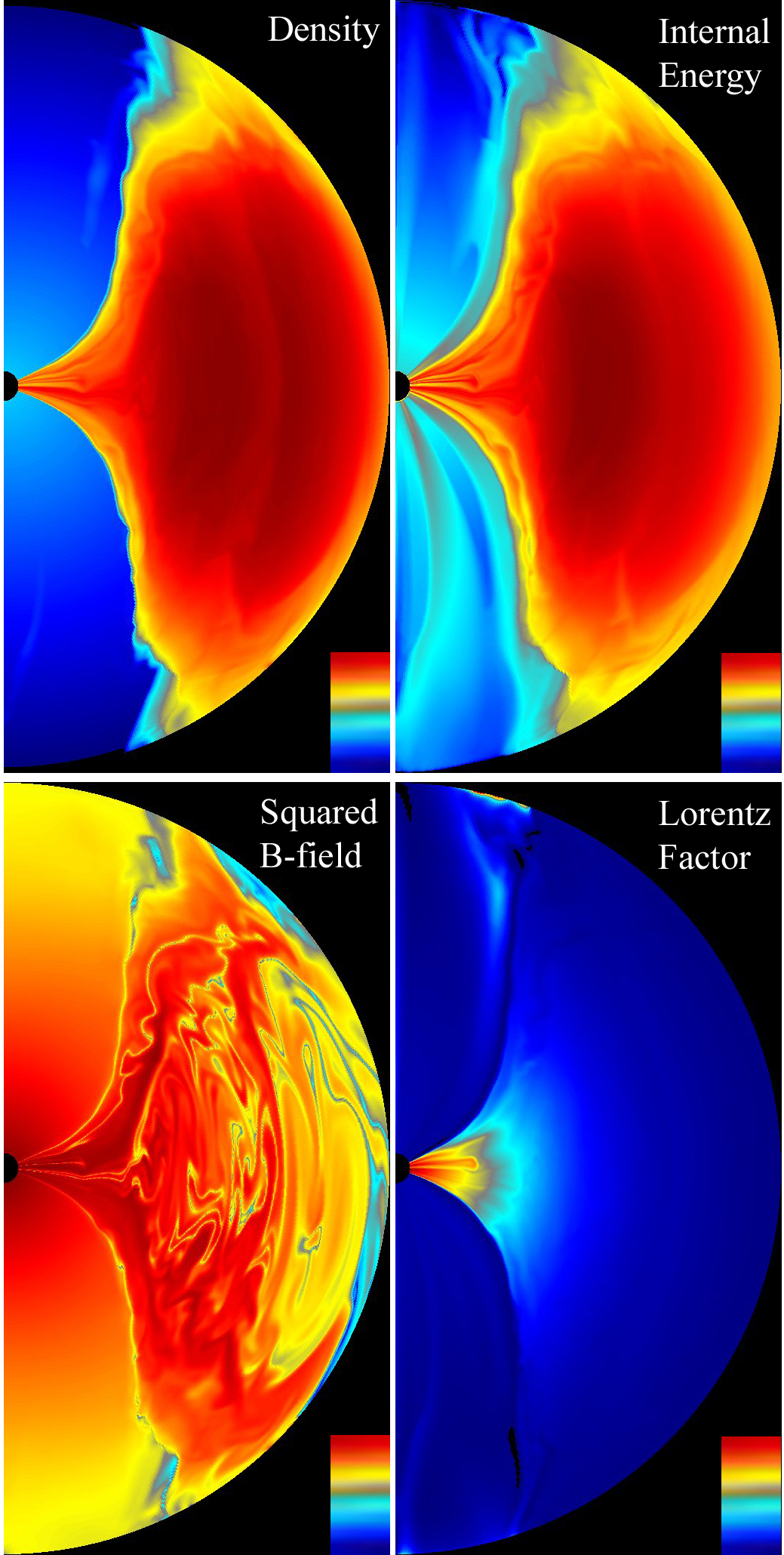}
  \caption{Panels from final timestep t = 8000 (1.4x10$^{5}$ s) of HARM data, as labelled for different physical quantities.  We show a cut-out in the  r-z plane, with the black hole's horizon shown on left.  As detailed in the text, the radial direction is scaled logarithmically, and angular dimension is scaled more finely toward the equator.  Therefore, the finest resolution is obtained closest to the horizon and equator.  This causes a ``spreading'' effect in the images, as they are shown evenly spaced by cell number, not physical value.  Each quantity is shown on a logarithmic scale, with dark red being the highest values, black being the lowest.}
\end{figure}

Our simulation was run to approximately 8000 timesteps, which equates to about 1.4x10$^{5}$ seconds, in physical time within the simulation.  Figure 1 displays and explains four panels, which show density, internal energy, squared magnetic field, and bulk lorentz factor through the simulation volume.  Of importance to note is that the images are shown evenly spaced in cell number.  This creates a kind of stretching effect near the equator and near the horizon -- causing the disk to appear much thicker than it actually is.  

As expected, the density concentrates itself around the equatorial plane, with the internal energy contours closely following.  The magnetic field is strongest in the disk region, but saturates the region, generally being stronger nearer the horizon.  More importantly, one can see how highly turbulent the field becomes after being seeded smoothly in the initial torus.  As the turbulence is expected to provide the mechanisms for electron heating, in future work, the level to which the field is churned up will be important to analyze.  The final panel shows the bulk gas lorentz factor, which is, as expected, closely related to distance from the black hole.  

\begin{figure}
  \includegraphics[scale=0.4]{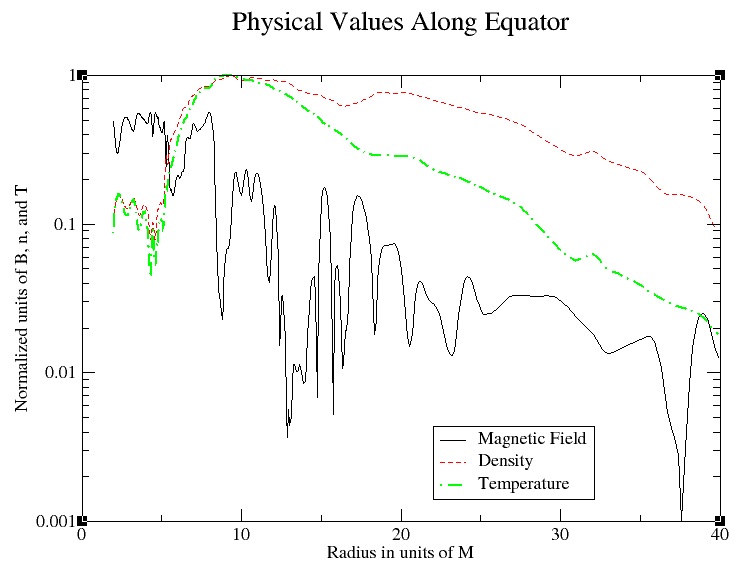}
  \caption{Profiles of magnetic field, density, temperature, along the equator, as a function of radius.  These quantities are normalized to the maximum value for each.}
\end{figure}

\begin{figure}
  \includegraphics[scale=0.4]{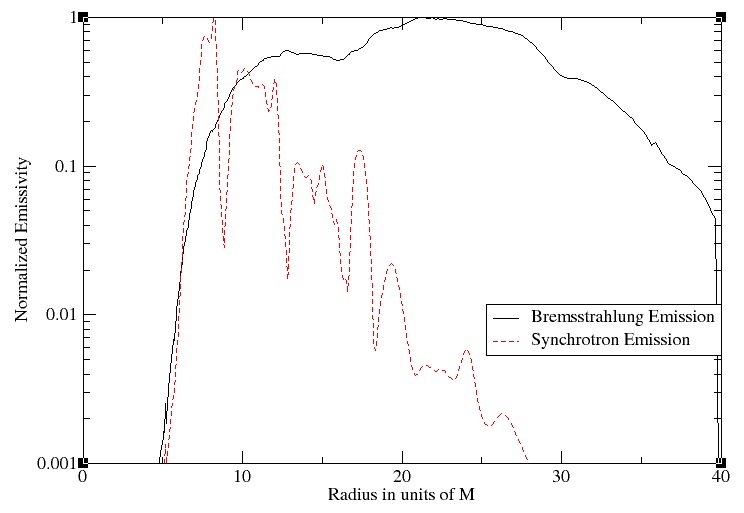}
  \caption{Profiles of optically thin synchrotron and bremsstrahlung emissivity.  These values are normalized to the maximum for each set, so no comparisons should be made between the strength of the two components from this plot.  These values are found by determining the emissivity within the closest cell and multiplying it by the radius and scale height at that radius.}
\end{figure}

\begin{figure}
  \includegraphics[scale=0.31]{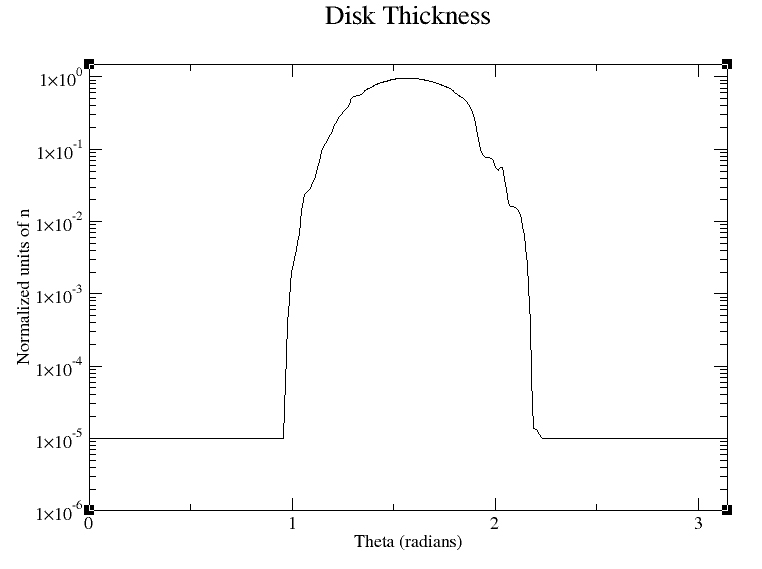}
  \caption{Profile of density along a constant radius r = 9 GM/c$^{2}$ slice over theta, from pole to pole.  For reference, at this radius, the scale height is 2.7 GM/c$^{2}$.}
\end{figure}

Figure 2 shows profiles of density, temperature, and magnetic field along an equatorial slice of the simulation volume.  Again of note is the high level of turbulence in the magnetic field.  Within 5 GM/c$^{2}$, the density and temperature drop to very low levels.  This corresponds well with the flow velocities as this is expected to be where the flow becomes almost completely radial, and may be mostly a product of how quickly matter falls to the horizon at this point.  Of some concern is how much these final profiles are due to the intial torus set-up.  Further trials would be needed to compare initial torus location to final results.  

Figure 3 also shows a profile of the synchrotron and bremsstrahlung emission as a function of radius.  To calculate this value, we multiplied the optically thin synchrotron emissivity and bremsstrahlung emissivity, respectively, by the radius and density scale height at that radius.  The synchrotron emissivity was found by averaging the magnetic field over a number of zones near the equator -- this helps eliminate some of the high variability seen in the magnetic field.  It should be noted that the overall shape did not change dramatically as this was averaged over a smaller or larger number of cells.  It is apparent that the majority of emission emerges between radii of approxiately 8M and 28M.  Areas inside and outside this region contribute decreasing amounts to the total emissivity.  It is important to note the lack of contribution from the 28M-40M region to overall emission, as our simulation volume for the MC code only extends to 28M (as is discussed below, in relation to the overlap of our two simulation grids).

To demonstrate actual disk thickness, density along a typical slice through a constant radius r = 9 GM/c$^{2}$ is shown in Figure 4.

\begin{figure}
  \includegraphics[scale=0.27]{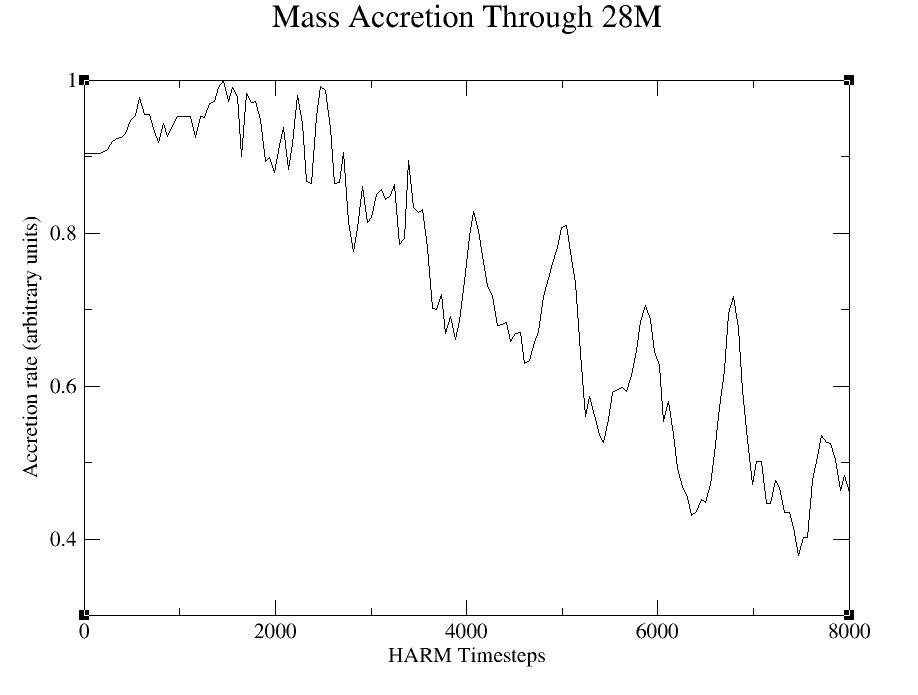}
  \caption{Mass and luminosity accretion rates onto the black hole through a surface at 28M as a function of HARM timestep -- with each timestep being equal to approximately 17 physical seconds.}
\end{figure}

Our last HARM output figure, Figure 5, shows the variance of mass accretion onto the black hole through a surface at 28M through time.  When comparing parameters required to generate spectra at quiescent and flaring data points, the amount that accretion varies could be important in determining the appropriateness of the variation in values used between the two points.  In general, at a late time, mass accretion varies up to a factor of two.  The time scale of accretion variation, on an order of several hours, is reasonable for the observed variations in Sgr A*'s spectrum.  As accretion variation is due not only to instabilities in the flow but also to changes in density of accreted matter, this factor of two is not considered a constraint on scaling.

\subsection{Spectrum determination with MC code}

The MC code used for photon emission and scattering has a long history and is discussed in a number of resources \citep{b8,b13,b7,b5,b31,b32,b33,b6}.  For a complete treatment, readers should see these papers.

In general, this code is a coupled MC/FP (Fokker-Planck) code.  For our intents, the FP evolution of the electron distribution was unnecessary at this stage, so it was turned off to allow a fixed temperature given by the HARM output.  The code is set up on a 2D axially-symmetric cylindrical grid, creating a (hollow or solid, depending on whether the inner radius is set to zero) cylindrical shape.  Each zone is assigned a density, ion and electron temperatures, magnetic field amplitude, and thermal and nonthermal distribution components.  For our purposes, this is simply set to be a Maxwellian, but the code allows power law nonthermal distributions as well.  The code allows emission from the volume and boundaries, and emitted photons are tracked and allowed to scatter or absorb.

This approach should be quite consistent, but it should be mentioned that the MC code does not include general relativistic effects.  Close to the horizon, the photons' paths should be bent by the high gravity -- but this code does not include this effect.  However, as was shown in the previous section, densities, temperatures, and emissivities are quite low, relatively, near the horizon.  Because of this, it is expected that these GR curvature effects would be minimally important for computing the global spectrum.

\subsubsection{Coupling GRMHD output to MC input}

\begin{figure}
  \begin{center}
  \includegraphics[scale=0.5]{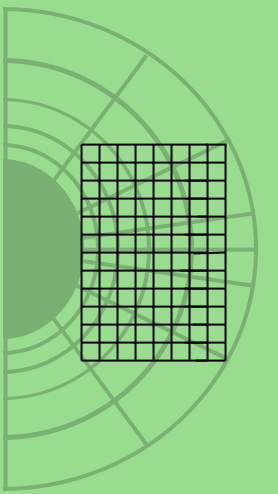}
  \caption{An exaggerated representation of the HARM and MC code grid overlays.  The spherically spaced grid underneath is that of the HARM code.  The rectangularly spaced grid overlaid is that of the MC code.  Both grids are axially symmetric, with said axis on the left side in this image.}
  \end{center}
\end{figure}

Shown in Figure 6 is an overlay of the grids for the two codes.  It can be easily seen that regardless of how our MC grid is set up, it will almost invariably undersample and oversample the HARM grid at different locations.  This is minimized as much as possible by making a very large MC grid -- in our trials, we use a 95x95 vertical by radial grid.  The grid can actually be quite a bit coarser than this before one notices a significant deviation in spectrum.  As we cannot cover the entire region with a rectangular grid, we have chosen a box which is 28 GM/c$^{2}$ radially and 56 GM/c$^{2}$ vertically.  This box is oriented so that its inner edge is at the horizon. 

As noted previously, the output from HARM is completely scalable by black hole mass.  For our trials, a rather well-accepted value of 3.6x10$^{6}$ solar masses was assigned.  We also have the freedom to choose maximum initial density and have the other parameters scale with the density consistently.  Density is scalable in this way since it only reflects a change in accretion rate.

In the HARM output, to determine temperature, we use internal energy.  As almost all contributions to this come from ions, we can determine an ion temperature in each zone, but not an electron temperature.  Coulomb coupling is very low due to the low densities expected in the accretion disk of Sagittarius A* (about 10$^{6}$ to 10$^{10}$ particles/cm$^{3}$), and as the nonthermal heating and cooling mechanisms are not well understood, maximum electron temperature is kept as a free parameter for spectral fitting purposes in this paper.  This means, in effect, we are setting a constant global ratio between T$_{i}$, the temperature of the ions, and T$_{e}$, the temperature of the electrons, and therefore assuming a two temperature flow, though this ratio is allowed to change between trials.

We consider the magnetic field given by the HARM data to be the MRI saturated field.  As mentioned previously, had we input an initial toroidal field, this would have contributed to the final total field.  For this reason, we use the MRI-saturated field as a lower limit to our field value and allow scaling above this.

So, in effect, we set a maximum density, which in turn sets an ion temperature and lower limit to the magnetic field.  The electron temperature is freely changed by changing the ratio between T$_{i}$ and T$_{e}$, and the magnetic field can be scaled up from its initial value as necessary.

\subsubsection{Trial and fitting procedure}

Figures 7-11 show observational data points for Sagittarius A*, whose distance is taken to be approximately 8 kpc, and trial fits.  Circles shown represent data points, upside down triangles are upper limit values, and the bowties are flaring and quiescent points from Chandra observations presented by \citet{b2}.

Data in the radio to IR range are fairly easy to fit on their own.  This is typically done with synchrotron emission arising from the acceleration of moving charges by a magnetic field.  When sufficiently energetic, these charges (electrons, in our case) produce a continuum spectrum whose flux and turnover frequency are directly related to the values of electron temperature, magnetic field, and density in this region.

In such a hot medium, synchrotron self-absorption becomes important.  This phenomenon occurs when a photon interacts with a charged particle in a magnetic field and transfers its energy to the particle.  The low energy positive slope of the synchrotron curve in each of the trials shown below is due to synchrotron self-absorption.

The difficulties arise in fitting the flaring or quiescent points in the x-ray simultaneously with the radio/IR data.  The approach made in this paper is to initially try to fit the flaring point.  Its flat slope in L$_{E}$ suggests that it might be fit by several possible components of the spectrum:  

\begin{enumerate}

	\item Bremsstrahlung emission, or free-free emission, arises (usually, and in this case) from the acceleration of a free electron by a free nucleus, in a completely ionized plasma.  This creates a relatively flat spectrum in L$_{E}$ out to a cut-off point which corresponds with the temperature of the plasma.  The flux of this component is directly related to the square of the density, as it depends on the population of electrons and ions, and also to the square root of the temperature.  Temperature also serves to locate the high energy cut-off in the spectrum.  Magnetic field values do not affect bremsstrahlung emission.

	\item Compton scattering is produced when photons interact with particles (in this case, electrons), leading to changes in energy for both photon and particle.  Of interest for our work is inverse Compton scattering, where a less energetic photon gains energy in an interaction with a hot electron.  In general, this tends to form a photon population whose shape is related to the initial photon distribution and the electron distribution it scatters from.  This spectrum is typically shifted up in energy an amount approximately equal to the square of the Lorentz factor of the electron population, and drops in flux by an amount equal to the optical depth of the scattering medium.  For our trials, the Compton spectrum of importance is mostly generated via synchrotron self-Comptonization (SSC), which refers to the Comptonization of a photon spectrum produced via synchrotron emission by the population of electrons responsible for emitting it to begin with.  As the final spectrum resembles the electron distribution and initial (unscattered) synchrotron spectrum, it is possible to locate this spectrum so its flat top intersects the flat flaring x-ray point.

	\item The second Compton scattering component, as its name suggests, is the spectral component created by the Compton scattering of an already scattered photon population.  In this case, it shares the characteristics of the first scattered spectrum, and therefore is a viable component to fit the flaring x-ray point.

\end{enumerate}

All information on emission processes can be found in \citet{b19}.

A large number of MC runs were done to find several sets of parameters which fit this point while remaining consistent with the order of magnitude estimate for total luminosity of 10$^{36}$ erg/s suggested by \citet{b22}.

The second step in fitting was to attempt to fit the quiescent x-ray data point by varying parameters from those found to fit the flaring point.  This was done, specifically, by using the same input values as the flaring point, but by dropping either the density or temperature until the spectrum went through the quiescent point.  Depending on the component being fit to the x-ray data, one or both of these parameters could be changed to lower the luminosity to intersect the quiescent point.  As we have data showing the variability of mass accretion by HARM, we can estimate reasonable changes in density.  Due to the unknown nature of electron heating, it is not inconsistent to suggest that flaring and quiescent points may also arise due to brief moments of higher or lower heating, caused by dissipation of large current sheets in the MRI-turbulence cascade.

It should be noted that observations show a strong correlation between x-ray and radio/IR flux during flares.  As it is generally accepted that the lower energy spectral component is due to synchrotron, this lends strength to the idea that the x-ray data is of synchrotron origin -- either as a self-Comptonization component, or an extended synchrotron component \citep{b39,b40,b41}.

\section{Results}

The results presented are shown first for fits to the flaring data (the higher x-ray point), then for fits to the quiescent data (the lower x-ray point).  The HARM data used in the MC runs are based on output at the last timestep (t=8000).  Fits are described by the component used to fit the x-ray data.  However, this does not indicate that other spectral components were turned off during these runs.  For instance, the bremsstrahlung spectrum can be seen to have both first and second Compton bumps, but the spectrum at the x-ray point is almost purely bremsstrahlung.

\subsection{Flaring results}

\begin{figure}
  \includegraphics[scale=0.29]{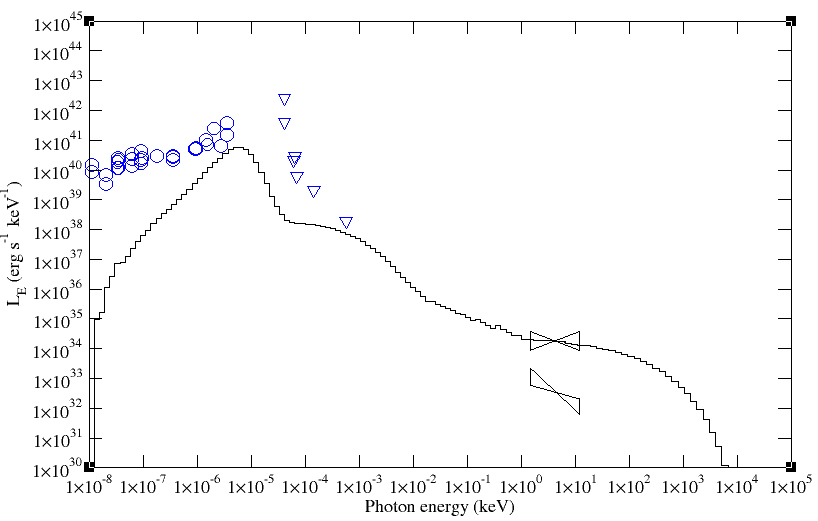}
  \caption{Fit to flaring data with bremsstrahlung component.  Maximum values of the 95x95 cell grid are scaled to n = 4.3x10$^{9}$ particles/cm$^{3}$, T = 1.2x10$^{3}$ keV, B = 9.9x10$^{2}$ G.  Total luminosity is 3.4x10$^{36}$ erg/s.}
\end{figure}

Figure 7 shows an attempt to fit the flaring data with the spectrum's bremsstrahlung component.  This is similar to fits attempted by \citet{b18}.  Like their trials, we have found that the bremsstralung component is too high when the radio data are fit well by the synchrotron data.  In this case, we have a simulation volume which extends to 28  GM/c$^{2}$, while \citet{b18} has volumes which extend to either 10 R$_{S}$ or 30 R$_{S}$ (20 GM/c$^{2}$ or 60  GM/c$^{2}$, respectively).  They found adequate fits with the smaller volume with a density maximum of 1x10$^{10}$ particles/cm$^{3}$ and with the larger volume at 3.4x10$^{8}$ particles/cm$^{3}$.  As would be expected, our trial lies somewhere between the two.  However, it seems unlikely that any skewing of density or temperature will allow a fit to the slope of the quiescent data.

\begin{figure}
  \includegraphics[scale=0.29]{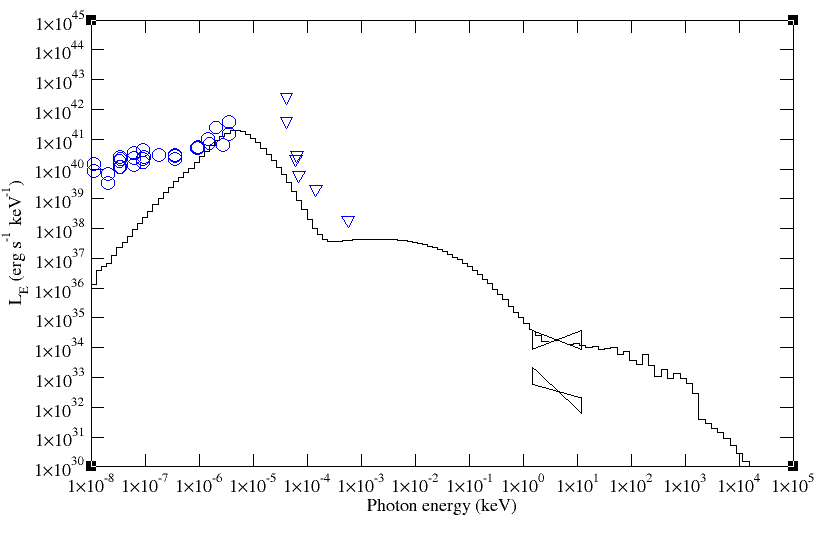}
  \caption{Fit to flaring data with second Compton bump component.  Maximum values of the 95x95 cell grid are scaled to n = 3.4x10$^{8}$ particles/cm$^{3}$, T = 8.2x10$^{3}$ keV, B = 1.6x10$^{2}$ G.  Total luminosity is 7.5x10$^{36}$ erg/s.}
\end{figure}

Figure 8 shows a fit by lowering density from the bremsstrahlung fit, but compensating by increasing temperature until the flat portion of the twice-scattered Compton bump fits the slope at the flaring point.  Total luminosity and the spectral slope at the flaring point suggest this is a good fit -- and the shape of the spectrum seems to allude to a possible fit to the quiescent point's slope if density is allowed to drop.  As with the bremsstrahlung fit, we can compare the density maximum used in this trial to those of \citet{b18}.  The value here correlates to that group's lower density and higher volume value -- this is reasonable, as we have a significantly higher temperature now.

\begin{figure}
  \includegraphics[scale=0.29]{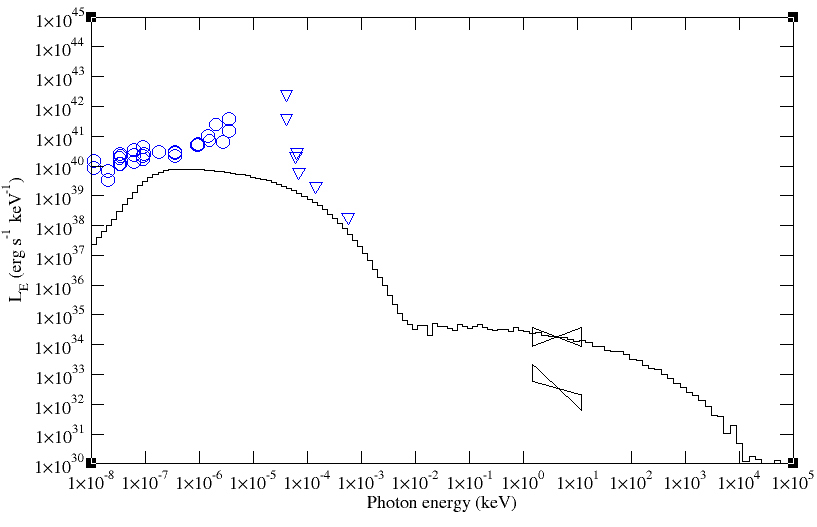}
  \caption{Fit to flaring data with first Compton bump component.  Maximum values of the 95x95 cell grid are scaled to n = 5.2x10$^{6}$ particles/cm$^{3}$, T = 1.8x10$^{5}$ keV, B = 13 G.  Total luminosity is 2.6x10$^{36}$ erg/s.}
\end{figure}

Figure 9 shows a fit with low density and high temperature, so the first Compton bump fits the flaring x-ray data point.  The synchrotron/synchrotron self-Compton approach was used analytically by \citet{b14}, which, while a one zone approximation, found very similar values to our maximums for one of their trials:  n = 2.3x10$^{7}$ particles/cm$^{3}$, T = 1.1x10$^{5}$ keV, B = 10.2 G.  This fit, like the one above, is promising due to its shape.  By lowering only temperature from this value, it appears the quiescent point may be fit with a similar slope.

\subsection{Quiescent results}

As alluded to above, quiescent fits for the bremsstrahlung trial, second Compton trial by varying the temperature, and first Compton trial by varying the density could be found which intersect the point.  However, all of these had nearly zero slopes in L$_{E}$, and were therefore unsatisfactory.  We present the two quiescent fits which appeared most promising.

\begin{figure}
  \includegraphics[scale=0.29]{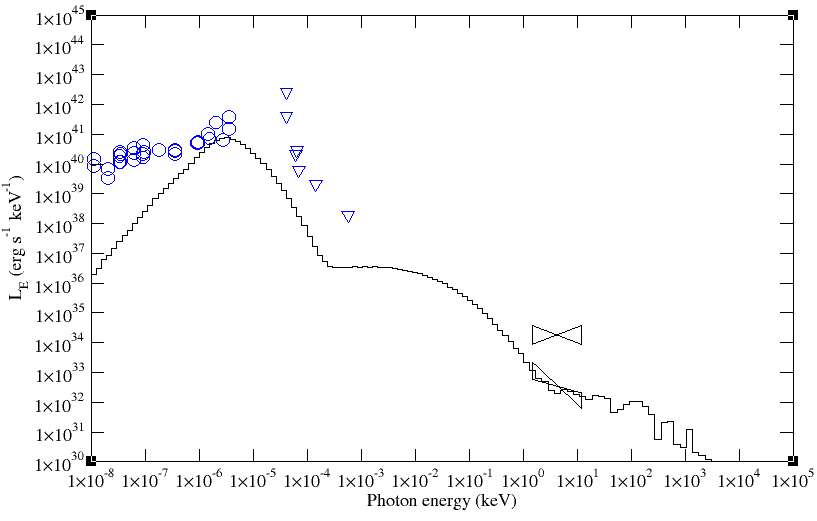}
  \caption{Fit to quiescent data with second Compton bump component.  Maximum values of the 95x95 cell grid are scaled to n = 6.9x10$^{7}$ particles/cm$^{3}$, T = 8.2x10$^{3}$ keV, B = 1.6x10$^{2}$ G.  Total luminosity is 8.2x10$^{35}$ erg/s.}
\end{figure}

Figure 10 shows a quiescent fit using the second Compton trial shown in Figure 8.  It should be noted, first of all, that the choppiness in this spectrum is due to lack of statistics in the twice-scattered photons.  Optical depth directly along the equator in this case is approximately 1x10$^{-3}$, leading to a very small population of scattered photons.  The slope at the quiescent point is significantly steeper than the previous trial at the flaring point, and fits well the slope required at quiescence.  In this case, we have dropped the density by less than a factor of five.  This is slightly higher than the accretion variation of HARM data, a factor of two, but is not extreme.  As noted in the variation analysis, variations in the density of accreted matter are not considered by HARM, and therefore, changes of this level are not precluded.  The results shown here can be changed greatly by small changes in input values, due to the combination of spectral components required to produce this spectrum.

\begin{figure}
  \includegraphics[scale=0.29]{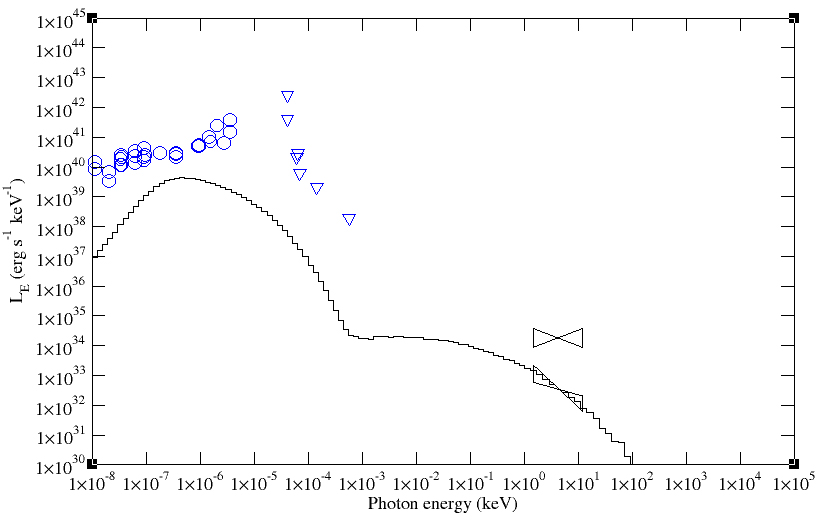}
  \caption{Fit to quiescent data with first Compton bump component.  Maximum values of the 95x95 cell grid are scaled to n = 5.2x10$^{6}$ particles/cm$^{3}$, T = 4.7x10$^{4}$ keV, B = 13 G.  Total luminosity is 3.4x10$^{34}$ erg/s.}
\end{figure}

Figure 11 shows the quiescent fit for the first Compton trial shown in Figure 8.  This fit was obtained by dropping the temperature by about half an order of magnitude.  As noted before, without better understanding of acceleration mechanisms, it is difficult to say whether global temperature changes of this scale are feasible or not.  However, by this trial, this approach looks promising, and is closely in line with the idea that correlated low and high energy flares arise due to synchrotron radiation -- in this case, the x-ray being fit by an SSC component.  This trial gives our best fit to the quiescent data point's slope.

\section{Conclusions}

We have presented results for coupled physical and spectral simulations of the accretion disk of Sagittarius A*.  Fits to flaring x-ray data are shown for three sets of parameters of an entirely thermal plasma, and these results are discussed, noting their similarities to two other recent models made for this same situation.  The values used in our trials for all three sets of parameters are quite similar (respectively) to those used by \citet{b14} and \citet{b18}.  

Of these, the trials which fit the flaring data point with the first Compton bump and second Compton bump show promise to allow fits to the quiescent x-ray data by only changing one physical value.  The second Compton bump trial shows a promising fit which intersects the quiescent point by dropping the density, and may be at an appropriate slope.  This trial is also the best fit to low energy data while fit to the x-ray points.  While intersecting each x-ray point, some of the IR data are fit by the synchrotron bump.  The first Compton bump trial shows a very good fit, both in slope and position, to the quiescent point by dropping the temperature, though these trials never match the IR or radio data points.  The second Compton bump fit appears approximately valid in comparison to the variations in mass accretion onto the black hole suggested by the HARM output.  It should be noted, as reported in \citet{b42}, the quiescent state X-ray emission can be attributed to thermal emission from the large scale accretion flow.  For this reason, the data point we have fit here could be considered an upper limit.

Overall, we present our second Compton bump trials as our best explanation for the observed spectra of Sagittarius A*.  Due to the combination of spectral components at the critical point in the x-ray, we are able to describe a large number of different slopes and locations -- with these being quite sensitive to input values.  This may account for the observed variability in x-ray slopes and luminosities.

While our second Compton trial can fit the IR points, it is found that none of our fits can match the radio data well while also fitting x-ray data -- this suggests that a completely thermal approach using GRMHD and one-temperature electrons is lacking in ways that may be improved by using FP techniques or results suggested by recent kinetic simulations of turbulence-heated electron distributions \citep{b38}.  It is also possible that a larger simulation volume must be enclosed to fit these points concurrently with the x-ray data.

In the future, this project will turn to some method of consistently introducing nonthermal particle acceleration to determine whether this can provide better fits to the joint radio/IR/x-ray data.

\section*{Acknowledgments}
This work was supported by a Los Alamos National Laboratory IGPP research grant and NSF-AST-0406882.

The authors would like to thank a number of people whose input helped shape this work, including, but not limited to, Markus B\"{o}ttcher, Xuhui Chen, Justin Finke, Charles Gammie, and Feng Yuan.

\label{lastpage}


\begin{thebibliography}{99}
\bibitem[\protect\citeauthoryear{Abramowicz et al.}{1995}]{b1} Abramowicz M. A., Chen X., Kato S., Lasota J.-P., Regev O. 1995, ApJ, 438, L37
\bibitem[\protect\citeauthoryear{Baganoff et al.}{2001}]{b2} Baganoff F. K., et al. 2001, Nature, 413, 45
\bibitem[\protect\citeauthoryear{Baganoff et al.}{2003}]{b3} Baganoff F. K., et al. 2003, ApJ, 591, 891
\bibitem[\protect\citeauthoryear{Balbus}{2003}]{b28} Balbus S. A. 2003, ARA\&A, 41, 555
\bibitem[\protect\citeauthoryear{Balick \& Brown}{1974}]{b4} Balick B., Brown R. 1974, ApJ, 194, 265
\bibitem[\protect\citeauthoryear{B\'{e}langer et al.}{2005}]{b37} B\'{e}langer G., Goldwurm A., Melia F., Ferrando P., Grosso N., Porquet N., Warwick R., Yusef-Zadeh F. 2005, ApJ, 635, 1095
\bibitem[\protect\citeauthoryear{B\"{o}ttcher et al.}{2003}]{b31} B\"{o}ttcher M., Jackson D. R., Liang E. P., 2003, ApJ, 586, 339
\bibitem[\protect\citeauthoryear{B\"{o}ttcher \& Liang}{1998}]{b5} B\"{o}ttcher M., Liang E. P.  1998, ApJ, 506, 281.
\bibitem[\protect\citeauthoryear{B\"{o}ttcher \& Liang}{2001}]{b6} B\"{o}ttcher M., Liang E. P.  2001, ApJ, 552, 248.
\bibitem[\protect\citeauthoryear{B\"{o}ttcher et al.}{1998}]{b7} B\"{o}ttcher M., Liang E. P., Smith I. A.  1998, A\&A, 339, 87.
\bibitem[\protect\citeauthoryear{Canfield et al.}{1987}]{b8} Canfield E., Howard W. M., Liang E. P.  1987, ApJ, 323, 565.
\bibitem[\protect\citeauthoryear{Dodds-Eden et al.}{2009}]{b41} Dodds-Eden et al. 2009, arXiv: 0903.3416
\bibitem[\protect\citeauthoryear{Eckart et al.}{2004}]{b35} Eckart A., et al. 2004, A\&A, 427, 1
\bibitem[\protect\citeauthoryear{Eckart et al.}{2006}]{b36} Eckart A., et al. 2006, A\&A, 450, 535
\bibitem[\protect\citeauthoryear{Finke}{2007}]{b33} Finke J., August 2007, Physics and Astronomy, Monte Carlo/Fokker-Planck simulations of Accretion Phenomena and Optical Spectra of BL Lacertae Objects
\bibitem[\protect\citeauthoryear{Finke \& B\"{o}ttcher}{2005}]{b32} Finke J. D., Bottcher M., 2005, PASP, 117, 483
\bibitem[\protect\citeauthoryear{Fishbone \& Moncrief}{1976}]{b9} Fishbone L. G., Moncrief V.  1976, ApJ, 207, 962.
\bibitem[\protect\citeauthoryear{Gammie et al.}{2003}]{b10} Gammie C. F., McKinney J. C., Toth G.  2003, ApJ, 589, 444
\bibitem[\protect\citeauthoryear{Ghez et al.}{2003}]{b11} Ghez A. M., et al. 2003, ApJ, 586, L127
\bibitem[\protect\citeauthoryear{Goldston et al.}{2005}]{b29} Goldston J. E., Quataert E., Igumenshchev I. V. 2005, ApJ, 621, 785
\bibitem[\protect\citeauthoryear{Hawley}{2001}]{b27} Hawley J. F. 2001, ApJ, 554, 534
\bibitem[\protect\citeauthoryear{Hawley \& Balbus}{2002}]{b12} Hawley J. F., Balbus S. A. 2002, ApJ, 573, 738
\bibitem[\protect\citeauthoryear{Ichimaru}{1977}]{b25} Ichimaru S. 1977, ApJ, 214, 840
\bibitem[\protect\citeauthoryear{Igumenshchev et al.}{2003}]{b30} Igumenshchev I. V., Narayan R., Abramowicz M. A. 2003, ApJ, 592, 1042
\bibitem[\protect\citeauthoryear{Kato et al.}{2004}]{b23} Kato Y., Mineshige S., Shibata K. 2004, ApJ, 605, 307
\bibitem[\protect\citeauthoryear{Liang}{2009}]{b38} Liang E., 2009, arXiv: 0902.4740
\bibitem[\protect\citeauthoryear{Liang \& Dermer}{1988}]{b13} Liang E. P., Dermer C. D.  1988, ApJ, 325, L39.
\bibitem[\protect\citeauthoryear{Liu \& Melia}{2001}]{b39} Liu S., Melia F. 2001, ApJ, 561, L77
\bibitem[\protect\citeauthoryear{Liu et al.}{2006}]{b14} Liu S., Petrosian V., Melia F., Fryer C. L.  2006, ApJ, 648, 1020.
\bibitem[\protect\citeauthoryear{Markoff et al.}{2001}]{b40} Markoff S., Falcke H., Yuan F., Biermann P. L. 2001, 379, L13
\bibitem[\protect\citeauthoryear{Martin et al.}{2009}]{b34} Martin K. W., Liu S., Fragile C., Yu C., Fryer C. L. 2009, arXiv: 0904.0118
\bibitem[\protect\citeauthoryear{Melia}{2006}]{b15} Melia F. 2006, The Galactic Supermassive Black Hole (Princeton: Princeton Univ. Press)
\bibitem[\protect\citeauthoryear{Narayan \& Yi}{1994}]{b16} Narayan R., Yi I. 1994, ApJ, 428, L13
\bibitem[\protect\citeauthoryear{Narayan et al.}{1995}]{b26} Narayan R., Yi I., Mahadevan R. 1995, Nature, 374, 623
\bibitem[\protect\citeauthoryear{Noble et al.}{2006}]{b17} Noble S. C., Gammie C. F., McKinney J. C., Del Zanna L.  2006, ApJ, 641, 626.
\bibitem[\protect\citeauthoryear{Ohsuga et al.}{2005}]{b18} Ohsuga K., Kato Y., Mineshige S.  2005, ApJ, 627, 782.
\bibitem[\protect\citeauthoryear{Rybicki \& Lightman}{1979}]{b19} Rybicki G. B., Lightman A. P. 1979, Radiative Processes in Astrophysics (New York: Wiley)
\bibitem[\protect\citeauthoryear{Sch\"{o}del et al.}{2007}]{b20} Sch\"{o}del R., et al.  2002, Nature, 419, 694.
\bibitem[\protect\citeauthoryear{Shakura \& Sunyaev}{1973}]{b24} Shakura N. I., Sunyaev R. A. 1973, A\&A, 24, 337
\bibitem[\protect\citeauthoryear{Xu et al.}{2006}]{b42} Xu Y., et al. 2006, ApJ, 640, 319
\bibitem[\protect\citeauthoryear{Yuan et al.}{2003}]{b21} Yuan F., Quataert E., Narayan R. 2003, ApJ, 598, 301
\bibitem[\protect\citeauthoryear{Yuan}{2007}]{b22} Yuan F.  2007, ASPC, 373, 95.
\end{thebibliography}
\end{document}